\documentstyle[aps]{revtex}

\begin{document}

\title{Quark mean field model with density dependent couplings
 for finite nuclei}

\author{Y. H. Tan$^2$, H. Shen$^{1,2,3}$,  P. Z. Ning$^{1,2,3}$}

\address{
$^{1}$CCAST(World Lab.), P.O.Box 8730, Beijing 100080, China\\
$^{2}$Department of Physics, Nankai University, 
      Tianjin 300071, China\\
$^{3}$Institute of Theoretical Physics, Beijing 100080, China\\
}

\maketitle

\begin{abstract}
The quark mean field model, which describes the
nucleon using the constituent quark model,
is applied to investigate the properties of finite nuclei.
The couplings of the scalar and vector mesons with quarks 
are made density dependent through direct coupling to the scalar field
so as to reproduce the relativistic 
Brueckner-Hartree-Fock results of nuclear matter. 
The present model provides satisfactory results on the properties 
of spherical nuclei, and predicts an increasing size of the nucleon 
as well as a reduction of the nucleon mass in the nuclear environment.

\vspace*{0.5cm}


\end{abstract}

\section{Introduction}

One of the most interesting topics in nuclear physics is to study the 
variations of hadron properties in nuclear medium.
Experimentally, the EMC effect reveals the medium modification of the 
internal structure of the nucleon~\cite{emc}.
The EMC effect motivated many theoretical works on the study of hadrons
in terms of quarks and gluons.  At present, we are still far away to
describe nucleons and nuclei in terms of quarks and gluons using 
quantum chromodynamics (QCD), which is believed to be the fundamental 
theory of strong interactions. Thus, it is desirable to build models 
which incorporate quark and gluon degrees of freedom and respect the
established theories based on hadronic degrees of freedom.
Such models are necessarily crude from various view points since
the study of the nuclear many-body systems on the fundamental 
level is intractable.

The quark-meson coupling (QMC) model proposed by Guichon~\cite{guichon}
provides a simple and attractive framework to incorporate quark
degrees of freedom in the study of nuclear many-body systems.
The model describes nuclear matter as nonoverlapping MIT bags 
in which the quarks interact self-consistently
with structureless scalar and vector mesons in the mean field 
approximation. The QMC model has been subsequently extended with 
reasonable success to various problems of nuclear matter and finite 
nuclei~\cite{qmc1,qmc2,qmc3,qmc4}.
The quark mean field (QMF) model as named in Ref.~\cite{qmf1} 
took the constituent quark model
for the nucleon instead of the MIT bag model, 
so that it could overcome the shortcoming of the QMC model 
as explained in Ref.~\cite{qmf2}.
The QMF model has been applied to study the properties of finite nuclei,
where the coupling constants were determined by the saturation properties 
of nuclear matter~\cite{qmf2}.

In this paper, we introduce density dependent couplings for the QMF
model, so that we can reproduce the relativistic 
Brueckner-Hartree-Fock (RBHF) results of nuclear matter~\cite{rbhf}.
The RBHF approach as a microscopic many-body theory based on 
hadronic degrees of freedom, provided very closely the nuclear
matter saturation properties, where the coupling constants and
the form factors of the one-boson exchange potential (OBEP)
were fixed from the nucleon-nucleon scattering and the deuteron
properties. The RBHF approach can be considered as parameter-free 
many-body theory, since all parameters in RBHF approach were 
adjusted to the two-nucleon problem at first.
Encouraged by the success of the RBHF approach in describing 
nuclear matter properties, there were many applications of the 
RBHF results on finite nuclei~\cite{ddrmf1,ddrmf2,ddrmf3,ddrmf4}. 
The relativistic mean field (RMF) 
theory with density dependent couplings was developed 
for finite nuclei, as well as reproducing the RBHF results of nuclear matter.
The RMF theory with density dependent couplings provided very good 
results on the nuclear properties
as the binding energies and the nuclear radii with including the
rearrangement contributions~\cite{ddrmf2,ddrmf3,ddrmf4}.
Both the RBHF and RMF approaches are based on hadronic degrees of freedom,
where the nucleons and mesons are treated as structureless particles.
Therefore, they are not applicable to study the quark effect in nuclei.

Hence, it is very interesting to connect the QMF model to the RBHF results
by density dependent couplings, so that we can investigate the 
medium modification of the nucleon structure and finite nuclei
properties, as well as reproducing the quantitative results of RBHF theory
for nuclear matter simultaneously. 
The consideration to introduce density dependent couplings of mesons with quarks
is based on the medium modifications of the meson properties.
There are many works discussing the variations of hadron 
properties~\cite{mes1,mes2,mes3,mes4}.
In the present model, the scalar mean field provides a strong attraction
and as a consequence the constituent quark mass is reduced in nuclear
medium. The mesons, which are actually composites of quarks, 
are influenced due to the presence of the scalar mean field.
Therefore, the density dependent couplings modeled through 
direct coupling to the scalar field represents, 
in a way, self-interaction in the scalar field.

The paper is organized as follows. In Sec. II we describe
the QMF model with density dependent couplings.
In Sec. III we present the parameters used in this paper,
and compare the nuclear matter results with RBHF results.
We then apply this model to finite nuclei and discuss the 
variations of the nucleon properties inside nuclei in Sec. IV. 
Section V is devoted to the summary.

\section{QUARK MEAN FIELD MODEL WITH DENSITY DEPENDENT COUPLING}

In this section, we give a brief description of the QMF model
with density dependent couplings. 
We follow the basic concept of the QMF model~\cite{qmf1,qmf2}, 
in which the couplings of the mesons with quarks are constants.
In the present work, we introduce density dependent couplings 
for the scalar ($\sigma$) and vector ($\omega$) mesons with quarks,
which are modeled as functions of the $\sigma$ mean field.
We include also the isovector meson, $\rho$, whose coupling with a quark
is assume to be a constant for simplicity. 
We note that the contribution from the $\rho$ meson is very weak for 
stable nuclei compared with those from the $\sigma$ and $\omega$ mesons.
In addition, the photon field should be included for finite system.
 
In the QMF model, the nucleon in nuclear medium is described 
in terms of the constituent quark model, in which the quarks 
couple with the $\sigma$, $\omega$, and $\rho$ mesons. 
We take the relativistic mean field (RMF) 
approximation~\cite{rmf1,rmf2,rmf3} for the meson fields. 
In the constituent quark model, the quarks in a nucleon satisfy the
following Dirac equation:
\begin{eqnarray}
\left[i\gamma_{\mu}\partial^{\mu}-m_q+g^q_{\sigma}(\sigma)\sigma
-g^q_{\omega}(\sigma)\omega\gamma^0
-g^q_{\rho}\rho\tau_3\gamma^0-\chi_c
\right] q(r) = 0.
\end{eqnarray}
where $m_q$ is the constituent quark mass and $\chi_c$ the confinement. 
The $g^q_{\sigma}(\sigma)$ and $g^q_{\omega}(\sigma)$ as functions of
the $\sigma$ mean field, are the couplings of the $\sigma$ and $\omega$
mesons with quarks, while the $g^q_{\rho}$ denotes the coupling constant 
of the $\rho$ meson with a quarks. 
We follow Ref.~\cite{qmf1,qmf2} to take into account the spin correlations
and remove the spurious center of mass motion, and then work out 
the effective nucleon mass $M_n^*$. 

For a static nuclear many-body system, the Lagrangian within the 
mean field approximation can be written as
\begin{eqnarray}
{\cal L}&=&\bar{\psi}\left[i\gamma_\mu\partial^\mu -
                 M_n^*(\sigma)-g_\omega(\sigma)\omega\gamma^0
                -g_\rho \rho \tau_3\gamma^0
                \right]\psi \\ \nonumber
&& -\frac{1}{2}\left[(\bigtriangledown\sigma)^2 +m_\sigma^2\sigma^2\right]
   +\frac{1}{2}\left[(\bigtriangledown\omega)^2 +m_\omega^2\omega^2\right]
   +\frac{1}{2}\left[(\bigtriangledown\rho)^2   +m_\rho^2\rho^2\right]
\\ \nonumber
&& -\bar{\psi}\left[\frac{e}{2}\left(1+\tau_3\right)\gamma^0 A \right]\psi 
   +\frac{1}{2}(\bigtriangledown A)^2.
\end{eqnarray}
Here, $\psi$ denotes the nucleon fields;
$\sigma$, $\omega$, and $\rho$ are the $\sigma$, $\omega$, and $\rho$ 
meson mean fields with masses $m_{\sigma}$, $m_{\omega}$, and $m_{\rho}$, respectively; $A$ denotes the photon field. 
The nuclear many-body system can be solved with the change 
of the nucleon properties under the influence of the scalar mean field, which is exclusively expressed 
in the effective nucleon mass, $M_n^*(\sigma)$, 
as a function of the $\sigma$ mean field.
The $\omega$ and $\rho$ mean fields do not cause any change of the nucleon properties~\cite{fleck}, and they appear merely as the energy shift. 
The relations between nucleon-meson couplings and 
quark-meson couplings have been given in Ref.~\cite{qmf2} 
as $g_{\omega}(\sigma)=3g_\omega^q(\sigma)$ and $g_\rho=g_\rho^q$.
We will apply the present model to study the properties of nuclear 
matter and finite nuclei in the following two sections.

\section{PARAMETERS AND NUCLEAR MATTER RESULTS}
In this section we will show the numerical results of nuclear matter
and discuss the density dependent couplings.
The change of the nucleon properties under the influence 
of the $\sigma$ mean field has been studied in Ref.~\cite{qmf2},
where the effective nucleon mass, $M_n^*$, has been shown as a 
function of the quark mass correction, $\delta m_q$.
Here $\delta m_q$ is related to $\sigma$ 
mean field as $\delta m_q=g_{\sigma}^q(\sigma)\sigma$.
We take the same quark mass and two types of confinement
($\chi_c=\frac{1}{2}kr^2$ and 
 $\chi_c=\frac{1}{2}kr^2 (1+\gamma^0)/2$ with $k=1000 \;\rm{MeV/fm^2}$) 
as those used in Ref.~\cite{qmf2}.

To perform the nuclear matter calculation, we take the Lagrangian 
given in Eq. (2), in which all the derivative terms of the meson fields
vanish for infinite nuclear matter due to the translational invariance.
Then, the nucleons and mesons obey the following Euler-Lagrange equations,
\begin{eqnarray}
[i\gamma_{\mu}\partial^{\mu} &-& M_n^*(\sigma)
     -g_\omega(\sigma)\omega\gamma^0
     -g_\rho\rho\tau_3\gamma^0]\psi
     = 0, \\ 
\nonumber\\
\nonumber
m_\sigma^2\; \sigma&=&-\frac{\partial M_n^*(\sigma)}{\partial\sigma}
                     \langle\bar\psi\psi\rangle
                    -\frac{\partial g_\omega(\sigma)}{\partial\sigma}\omega 
                     \langle\bar\psi\gamma^0\psi\rangle \\ 
 &=-&\frac{\partial M_n^*(\delta m_q)}{\partial\delta m_q} g_\sigma^q(\sigma)
    \langle\bar\psi\psi\rangle
   -\frac{\partial M_n^*(\delta m_q)}{\partial\delta m_q} 
          \frac{\partial g^q_\sigma(\sigma)}{\partial\sigma}\sigma 
                \langle\bar\psi\psi\rangle
          -3 \frac{\partial g^q_\omega(\sigma)}{\partial\sigma}\omega 
                \langle\bar\psi\gamma^0\psi\rangle ,\\
\nonumber \\
 m_\omega^2 \;\omega &=&
    g_\omega(\sigma) \langle \bar \psi \gamma^0 \psi \rangle=
3 g^q_\omega(\sigma) \langle \bar \psi \gamma^0 \psi \rangle, \\
\nonumber\\
 m_\rho^2 \;\rho &=&
g_\rho   \langle \bar \psi \tau_3\gamma^0 \psi \rangle=
g^q_\rho \langle \bar \psi \tau_3\gamma^0 \psi \rangle.
\end{eqnarray}
Here, the bracket $\langle$ $\rangle$ means the expectation value
of the operator between the nuclear ground state.
We note that the appearance of the last two terms in Eq. (4) is due to the variable couplings, which is similar to the inclusion of the
rearrangement terms in the density dependent RMF approach
based on hadronic degrees of freedom~\cite{ddrmf2}.   
The effective nucleon mass $M_n^*$ and its derivative with respect to
the $\delta m_q$ has been studied in Ref.~\cite{qmf2}.
We take the meson masses as $m_\sigma=550 \;\rm{MeV}$, 
$m_\omega=783 \;\rm{MeV}$, and $m_\rho =770 \;\rm{MeV}$.

We introduce density dependent couplings, 
$g^q_\sigma(\sigma)$ and $g^q_\omega(\sigma)$,
in the present model.
This consideration is based on the variation of
meson properities as the nuclear environment changes.
The mesons, which are actually composites of quarks,
are influenced by the $\sigma$ mean field~\cite{mes3,mes4}. 
The quark-meson couplings in nuclear medium may be 
modified as density increases. It is therefore reasonable
to attempt density dependent couplings, which contain
the complicated medium modification effectively.
This treatment is similar in spirit to the density dependent 
RMF approach based on hadronic degrees of freedom, which provided a 
realistic description of bulk properties of finite nuclei 
and nuclear matter~\cite{ddrmf1,ddrmf2,ddrmf3,ddrmf4}. 
The nucleon-meson couplings in the density dependent RMF approach
were found to decrease with increasing nuclear matter density.
This suggests a decreasing quark-meson couplings due to 
the relations between the nucleon-meson couplings with the 
quark-meson couplings.

we model the density dependent couplings, $g^q_\sigma$ and $g^q_\omega$, through direct couplings to the $\sigma$ mean field.
We parameterize these variable couplings as
\begin{eqnarray}
  g^q_\sigma(\sigma)&=&a_\sigma-b_\sigma\sigma^{1/3}, \\
  g^q_\omega(\sigma)&=&a_\omega-b_\omega \sigma ^{1/3}.
\end{eqnarray}
With these formulas, it is possible to reproduce the scalar and vector 
potentials of the RBHF theory for symmetric nuclear matter
within a good accuracy. We list the parameters of $a_i$ and $b_i$ in Table I, which are determined by the RBHF results with potential A~\cite{rbhf}. 
The parameter $g^q_\rho$, which is still a constant for simplicity, 
is taken as $g^q_\rho=4$, so that we can get the symmetry 
energy to be around $35 \;\rm{MeV}$. 

In Fig. 1, we plot the scalar and
vector potentials, $U_S$ and $U_V$, as functions of the nuclear matter density $\rho$. The RBHF results~\cite{rbhf} are marked by solid dots, 
as well as the results in the RMF model with TM1 parameter set~\cite{rmf3}
by dotted curve for comparison. We show in Fig. 2 the energy per nucleon, $E/A$, as a function of density $\rho$.
The solid and dashed curves, which are the results of the 
present model with two types of confinement, are almost identical.
We note that the saturation density, binding energy and incompressibility
of the solid curve in Fig. 2 are $0.16 \;\rm{fm^{-3}}$, 
$-15.5 \;\rm{MeV}$ and  $159 \;\rm{MeV}$, respectively.
We found there exist slight differences between
the RBHF results and those in the present model in Fig. 2, 
which is partly due to the appearance of the rearrangement 
terms in Eq. (4). The importance of rearrangement terms
in a relativistic density dependent field theory
has been discussed extensively in Ref.~\cite{ddrmf2}.
   
\section{FINITE NUCLEI RESULTS}
In this section we will present the calculated results of some
spherical nuclei and discuss the variations of the 
nucleon properties inside nuclei.
The meson mean fields in a spherically symmetric nucleus 
are functions of the radial coordinate $r$.  
Here, we follow the prescription of Ref.~\cite{qmc3,qmf2}
and assume the nucleons obey the Dirac equation,
\begin{eqnarray}
\left[i\gamma_{\mu}\partial^{\mu} - M_n^*(\sigma)
     -g_\omega(\sigma)\omega(r)\gamma^0
     -g_\rho\rho(r)\tau_3\gamma^0
       -\frac{e}{2}(1+\tau_3) A(r)\gamma^0
\right]\psi
    = 0, 
\end{eqnarray}
where the effective nucleon mass $M^*_n$ is a function of the 
scalar mean field $\sigma$. 
The meson mean fields ($\sigma$, $\omega$, $\rho$) and the photon
field ($A$) satisfy the following Klein-Gordon equations, 
\begin{eqnarray}
\nonumber
\Delta\sigma(r)-m_\sigma^2 \sigma(r) &=&
\frac{\partial M_n^*(\sigma)}{\partial\sigma}
                     \langle\bar\psi\psi\rangle
+\frac{\partial g_\omega(\sigma)}{\partial\sigma}\omega(r) 
                     \langle\bar\psi\gamma^0\psi\rangle \\ 
&=&\frac{\partial M_n^*(\delta m_q)}{\partial\delta m_q} g_\sigma^q(\sigma)
                     \langle\bar\psi\psi\rangle
  +\frac{\partial M_n^*(\delta m_q)}{\partial\delta m_q} 
          \frac{\partial g^q_\sigma(\sigma)}{\partial\sigma}\sigma(r) 
                \langle\bar\psi\psi\rangle
          +3\frac{\partial g^q_\omega(\sigma)}{\partial\sigma}\omega(r) 
                \langle\bar\psi\gamma^0\psi\rangle ,\\
\nonumber \\
\Delta\omega(r)-m_\omega^2 \omega(r) &=&
    -g_\omega(\sigma) \langle \bar \psi \gamma^0 \psi \rangle=
-3 g^q_\omega(\sigma) \langle \bar \psi \gamma^0 \psi \rangle, \\
\nonumber\\
\Delta\rho(r)-m_\rho^2 \rho(r) &=&
-g_\rho   \langle \bar \psi \tau_3\gamma^0 \psi \rangle=
-g^q_\rho \langle \bar \psi \tau_3\gamma^0 \psi \rangle, \\
\nonumber\\
\Delta A(r) &=&
-\frac{e}{2}\langle \bar \psi (1+\tau_3)\gamma^0 \psi \rangle.
\end{eqnarray}

We solve the above equations
self-consistently using the parameters determined
by the RBHF results of nuclear matter with potential $A$.
In Fig. 3 and 4, we plot the charge density distributions for 
$^{40}\rm{Ca}$ and $^{208}\rm{Pb}$, respectively.
The experimental values taken from Ref.~\cite{exp1} are shown by solid curves.
The calculated results with scalar confining potential 
($\chi_c=\frac{1}{2}kr^2$) are shown by dashed curves, while 
those with scalar-vector confining potential 
($\chi_c=\frac{1}{2}kr^2 (1+\gamma^0)/2$) by dash-dotted curves.
The curves with two types of confinement are very close, since the 
parameters used here are determined by the same RBHF results.
The theoretical central densities are slightly higher than 
the experimental values, which is related to the RBHF results used 
in determining the parameters.
The dependence of the finite nuclei properties on the RBHF results
were discussed in more detail in Refs.~\cite{ddrmf1,ddrmf3}.   
The calculated results in the present model 
for the binding energies per nucleon $E/A$,
the rms charge radii $R_c$, and the spin-orbit spitting 
$\triangle E_{LS}$ are compared with the experimental values in Table II.
The experimental data taken from Refs.~\cite{exp1,exp2} are listed in the 
last column. The results with two types of confinement agree very well
with the experiment, and no explicit difference between them. 
In Table III and IV, the single particle energies 
of $^{40}\rm{Ca}$ and $^{208}\rm{Pb}$ are compared with 
the experimental data~\cite{exp2}. 
We show in Fig. 5 and 6 the scalar and vector mean fields
($\sigma$ and $\omega$) as functions of $r$ for $^{40}\rm{Ca}$ and $^{208}\rm{Pb}$. Here, the results with $\chi_c=\frac{1}{2}kr^2$
are given by solid curves, while those with 
$\chi_c=\frac{1}{2}kr^2 (1+\gamma^0)/2$ by dashed curves.
Again, the corresponding results with two types of confinement 
are very similar due to the same reason as mentioned before.

It is very interesting to see the variations of the nucleon properties
inside nuclei. We plot in Fig. 7 and 8 the ratios of the nucleon 
rms radius and effective mass in $^{40}\rm{Ca}$ and 
$^{208}\rm{Pb}$ to those in free space, $R/R_0$ and $M^*_n/M_n$,
as functions of nuclear radial coordinate $r$.
We find the model with $\chi_c=\frac{1}{2}kr^2 (1+\gamma^0)/2$ 
(dashed curves) gives near $10\%$ increasing nucleon radius
at the center of the nuclei,
while that with $\chi_c=\frac{1}{2}kr^2$ gives around $5\%$ increase.
The nucleon radius decrease to its value in free space
from center to outside, as the scalar mean field decreasing.
As for the effective nucleon mass, both scalar and scalar-vector
confinement cases give around $40\%$ reduction at the center of the nuclei.
The curves with two types of confinement are almost identical. 
It is due to the reproduction of the RBHF scalar potential.

\section{CONCLUSION}
We have developed the QMF model with density dependent
couplings, so that we can investigate the variations of 
the nucleon structure and finite nuclei properties, 
as well as reproducing the RBHF results of nuclear matter 
simultaneously. 
We introduced the density dependence of the couplings
of mesons with quarks through direct couplings to the $\sigma$ field
based on the capability of $\sigma$ field
in influencing hadron properties.
We have used the constituent quark model for the nucleon, 
which naturally allows the direct coupling of $\sigma, \omega$, and
$\rho$ mesons with quarks. With the parameters determined
by the RBHF results of nuclear matter, we can perform the numerical 
calculations for finite nuclei.

We have found the QMF model with density dependent couplings 
gave very successful descriptions of spherical nuclei.
By comparison with the density dependent RMF approach on hadronic level,
we have reproduced the properties of finite nuclei with the same 
or even better agreement to the experimental data. 
The present model with scalar confinement has the capability
to produced better results on the binding energies per nucleon 
compared with those in Ref.~\cite{qmf2}.  
Very weak dependence on the type of the confinement was found in
the results of nuclear properties, which is due to the reproduction 
of the RBHF scalar and vector potentials for nuclear matter.

We have also investigated the variations of the nucleon properties
in spherical nuclei. The QMF model with density dependent couplings
can provide a significantly swollen nucleon radius in nuclear environment.
We found the model with $\chi_c=\frac{1}{2}kr^2 (1+\gamma^0)/2$ 
predicted near $10\%$ increasing nucleon radius at the center of nuclei,
while that with $\chi_c=\frac{1}{2}kr^2$ gave around $5\%$ increase.
At the same time, around $40\%$ reduction of the nucleon mass at the 
center of nuclei were predicted.

\section*{Acknowledgments}
One of the authors (H.S.) would like to thank H. Toki for valuable
discussions. This work was supported in part by the National Natural 
Science Foundation of China.


\newpage
\begin{table}
\caption{Parameters needed to reproduce the RBHF results of 
         nuclear matter with potential $A$ in 
         Ref.~\protect\cite{rbhf} }
\vspace{0.5cm}
\begin{tabular}{lcccccc}
  & & QMF & &QMF & & \\
  & &$(\chi_c =\frac{1}{2}kr^2) $& 
    &$(\chi_c=\frac{1}{2}kr^2(1+\gamma^0)/2)$& & \\
\hline
$a_\sigma$                        & &  6.415 & & 7.328 & & \\
$b_\sigma$($\rm{fm^{1/3}}$)             & &  5.852 & & 6.016 & & \\
$a_\omega$                        & &  7.543 & & 7.543 & & \\
$b_\omega$($\rm{fm}^{1/3}$)             & &  6.470 & & 6.470 & & \\
\end{tabular}
\end{table}

\begin{table}
\caption{Binding energies per nucleon ($E/A$), rms charge radii ($R_c$),
and spin-orbit splittings for neutrons ($\triangle E_{n}$)
and protons ($\triangle E_{p}$) are compared with the 
experimental data~\protect\cite{exp1,exp2}.
The energies are in $\rm{MeV}$, and the radii are in $\rm{fm}$. }
\vspace{0.5cm}
\begin{tabular}{lclcccccc}
& & & & QMF & & QMF & & Expt. \\
& & & & $(\chi_c=\frac{1}{2}kr^2)$ & 
      & $(\chi_c=\frac{1}{2}kr^2(1+\gamma^0)/2)$ & &  \\
\hline
$^{40}$\rm{Ca} 
& & $E/A$              & & 8.65 & & 8.62 & & 8.55 \\
& & $R_c$              & & 3.46 & & 3.45 & & 3.45 \\
& & $\triangle E_{n}
 (1d_{5/2}-1d_{3/2})$  & & 4.5  & & 4.5  & & 6.3  \\
& & $\triangle E_{p}
 (1d_{5/2}-1d_{3/2})$  & & 4.4  & & 4.5  & & 7.2  \\
$^{48}$\rm{Ca} 
& & $E/A$              & & 8.58 & & 8.55 & & 8.67 \\
& & $R_c$              & & 3.49 & & 3.48 & & 3.45 \\
& & $\triangle E_{n}
 (1d_{5/2}-1d_{3/2})$  & & 3.8  & & 3.8  & & 3.6  \\
& & $\triangle E_{p}
 (1d_{5/2}-1d_{3/2})$  & & 3.9  & & 3.9  & & 4.3  \\
$^{90}$\rm{Zr} 
& & $E/A$              & & 8.50 & & 8.49 & & 8.71\\
& & $R_c$              & & 4.29 & & 4.28 & & 4.26 \\
& & $\triangle E_{n}
 (2p_{3/2}-2p_{1/2})$  & & 1.2  & & 1.3  & & 0.5  \\
& & $\triangle E_{p}
 (2p_{3/2}-2p_{1/2})$  & & 1.2  & & 1.2  & &      \\
$^{208}$\rm{Pb} 
& & $E/A$              & & 7.62 & & 7.62 & & 7.87 \\
& & $R_c$              & & 5.57 & & 5.57 & & 5.50\\
& & $\triangle E_{n}
 (2f_{7/2}-2f_{5/2})$  & & 1.6  & & 1.6  & & 1.8 \\
& & $\triangle E_{p}
 (1g_{9/2}-1g_{7/2})$  & & 2.6  & & 2.5  & & 4.0\\
\end{tabular}
\end{table}

\begin{table}
\caption{Single particle energies of neutron (n) and proton (p) 
for $^{40}\rm{Ca}$. The experimental data
are taken from Ref.~\protect\cite{exp2}. 
All energies are in $\rm{MeV}$.}
\begin{tabular}{lllccccl}
& & & QMF & & QMF & & Expt. \\
& & & $(\chi_c=\frac{1}{2}kr^2)$ & & 
$(\chi_c=\frac{1}{2}kr^2(1+\gamma^0)/2)$ & &  \\
&Shell& &n\hspace{1cm}p& &n\hspace{1cm}p& &n\hspace{1.5cm}p \\
\hline
&$ 1s_{1/2}  $& &51.3\hspace{1cm}43.1& &51.5\hspace{1cm}43.3
                         & &50.0\hspace{1cm}50$\pm$11\\
&$ 1p_{3/2}  $& &36.2\hspace{1cm}28.4& &36.3\hspace{1cm}28.5
                         & &30.0\hspace{1cm}34$\pm$6\\
&$ 1p_{1/2}  $& &33.6\hspace{1cm}25.7& &33.7\hspace{1cm}25.8
                         & &27.0\hspace{1cm}34$\pm$6\\
&$ 1d_{5/2}  $& &21.4\hspace{1cm}14.0& &21.5\hspace{1cm}14.0
                         & &21.9\hspace{1cm}15.5 \\
&$ 1d_{3/2}  $& &16.9\hspace{1cm} 9.5& &17.0\hspace{1cm} 9.6
                         & &15.6\hspace{1cm} 8.3 \\
&$ 2s_{1/2}  $& &17.3\hspace{1cm}10.0& &17.3\hspace{1cm}10.0
                         & &18.2\hspace{1cm}10.9 \\
\end{tabular}
\end{table}

\begin{table}
\caption{Single particle energies of neutron (n) and proton (p) 
for $^{208}\rm{Pb}$. The experimental data
are taken from Ref.~\protect\cite{exp2}. 
All energies are in $\rm{MeV}$.}
\begin{tabular}{lllccccl}
& & & QMF & & QMF & & Expt. \\
& & & $(\chi_c=\frac{1}{2}kr^2)$ & & 
$(\chi_c=\frac{1}{2}kr^2(1+\gamma^0)/2)$ & &  \\
&Shell& &n\hspace{1cm}p& &n\hspace{1cm}p& &n\hspace{1.5cm}p \\
\hline
&$ 1s_{1/2}  $& &56.3\hspace{1cm}43.4& &56.1\hspace{1cm}43.4
                         & &    \hspace{1cm}     \\
&$ 1p_{3/2}  $& &50.2\hspace{1cm}38.0& &50.1\hspace{1cm}37.9
                         & &    \hspace{1cm}     \\
&$ 1p_{1/2}  $& &49.9\hspace{1cm}37.6& &49.8\hspace{1cm}37.5    
                         & &    \hspace{1cm}     \\
&$ 1d_{5/2}  $& &43.1\hspace{1cm}31.3& &43.0\hspace{1cm}31.3
                         & &    \hspace{1cm}     \\
&$ 1d_{3/2}  $& &42.3\hspace{1cm}30.4& &42.2\hspace{1cm}30.3
                         & &    \hspace{1cm}     \\
&$ 2s_{1/2}  $& &39.7\hspace{1cm}27.3& &39.7\hspace{1cm}27.3
                         & &    \hspace{1cm}     \\
&$ 1f_{7/2}  $& &35.2\hspace{1cm}23.7& &35.1\hspace{1cm}23.7
                         & &    \hspace{1cm}     \\
&$ 1f_{5/2}  $& &33.6\hspace{1cm}22.1& &33.6\hspace{1cm}22.1
                         & &    \hspace{1cm}     \\
&$ 2p_{3/2}  $& &30.2\hspace{1cm}18.2& &30.2\hspace{1cm}18.2
                         & &    \hspace{1cm}     \\
&$ 2p_{1/2}  $& &29.6\hspace{1cm}17.5& &29.6\hspace{1cm}17.5
                         & &    \hspace{1cm}     \\
&$ 1g_{9/2}  $& &26.6\hspace{1cm}15.5& &26.6\hspace{1cm}15.5
                         & &    \hspace{1.6cm}15.4 \\
&$ 1g_{7/2}  $& &24.2\hspace{1cm}13.0& &24.2\hspace{1cm}13.0
                         & &    \hspace{1.6cm}11.4 \\
&$ 2d_{5/2}  $& &20.9\hspace{1cm} 9.1& &20.9\hspace{1cm} 9.2
                         & &    \hspace{1.6cm} 9.7 \\
&$ 2d_{3/2}  $& &19.7\hspace{1cm} 7.9& &19.7\hspace{1cm} 8.0
                         & &    \hspace{1.6cm} 8.4 \\
&$ 3s_{1/2}  $& &18.9\hspace{1cm} 6.8& &18.9\hspace{1cm} 6.9
                         & &    \hspace{1.6cm} 8.0 \\
&$ 1h_{11/2} $& &17.8\hspace{1cm} 6.9& &17.8\hspace{1cm} 7.0
                         & &    \hspace{1.6cm} 9.4 \\
&$ 1h_{9/2}  $& &14.4\hspace{1.6cm}    & &14.4\hspace{1.6cm}
                         & &10.8\hspace{1.6cm}     \\
&$ 2f_{7/2}  $& &11.9\hspace{1.6cm}    & &11.9\hspace{1.6cm}
                         & & 9.7\hspace{1.6cm}     \\
&$ 2f_{5/2}  $& &10.3\hspace{1.6cm}    & &10.3\hspace{1.6cm}
                         & & 7.9\hspace{1.6cm}     \\
&$ 3p_{3/2}  $& & 9.6\hspace{1.6cm}    & & 9.6\hspace{1.6cm}
                         & & 8.3\hspace{1.6cm}     \\
&$ 3p_{1/2}  $& & 9.0\hspace{1.6cm}    & & 8.9\hspace{1.6cm}
                         & & 7.4\hspace{1.6cm}     \\
&$ 1i_{13/2} $& & 8.9\hspace{1.6cm}    & & 8.9\hspace{1.6cm}
                         & & 9.0\hspace{1.6cm}     \\
\end{tabular}
\end{table}

\newpage
\section*{Figure captions}

\begin{description}

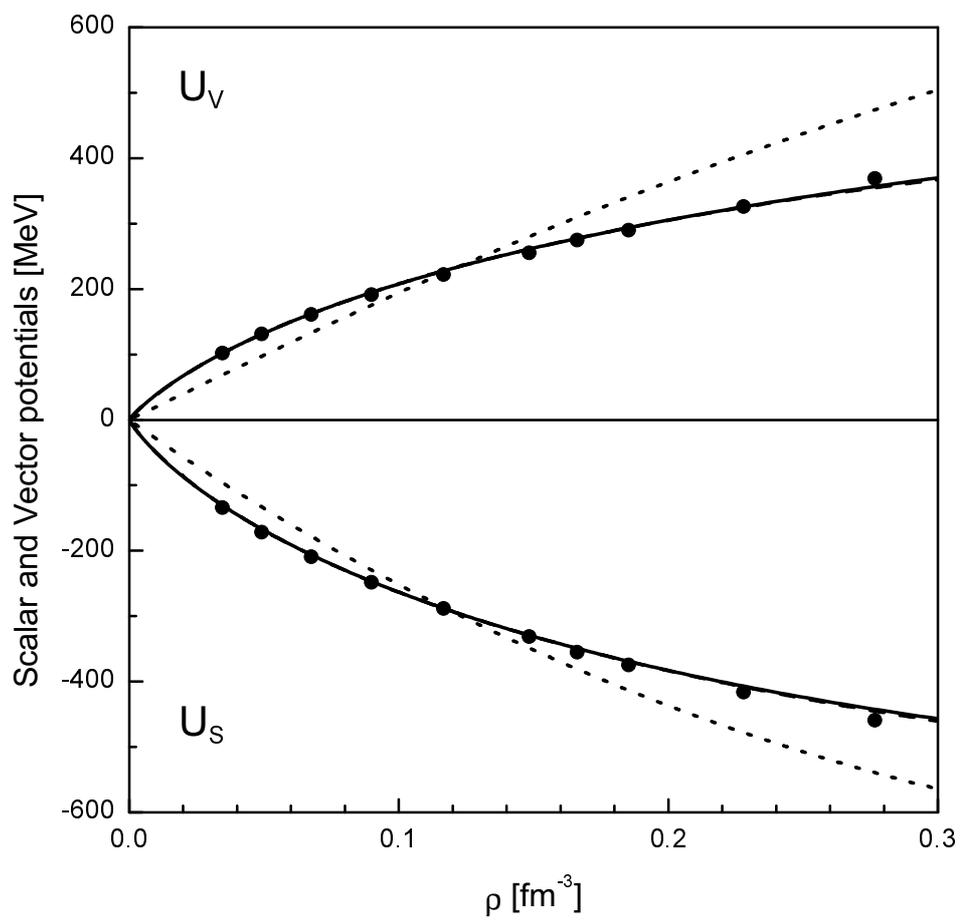
\item[Figure 1:]
The scalar and vector potentials of nuclear matter, 
$U_S$ and $U_V$, as functions of the nuclear matter density $\rho$.  
The results of the present model with $\chi_c=\frac{1}{2}kr^2$ 
are shown by solid curves, while those with $\chi_c=\frac{1}{2}kr^2(1+\gamma^0)/2$ are shown by dashed curves.  
The RBHF results with potential $A$ in Ref.~\protect\cite{rbhf}
are marked by solid dots, and the results in the RMF(TM1) 
model~\protect\cite{rmf3} are plotted by dotted curves for comparison.

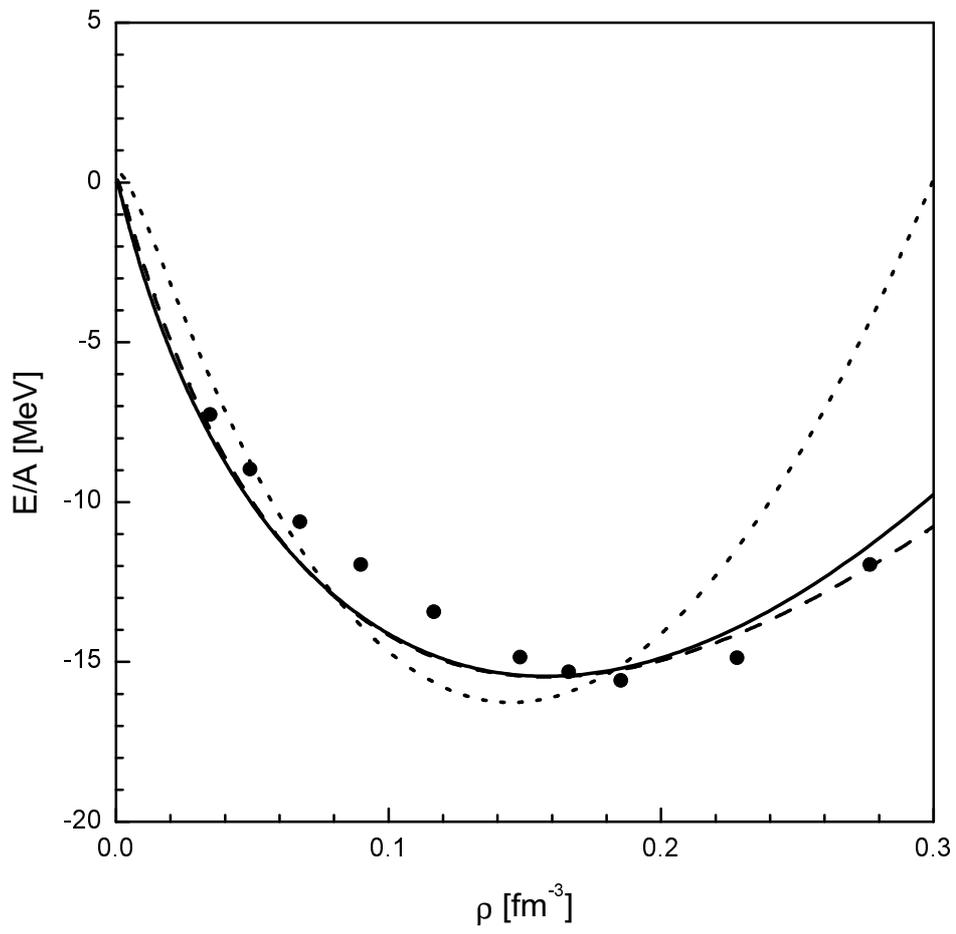
\item[Figure 2:]
The energy per nucleon, $E/A$, as functions of
the nuclear matter density $\rho$. The curves are labeled as in Fig. 1.

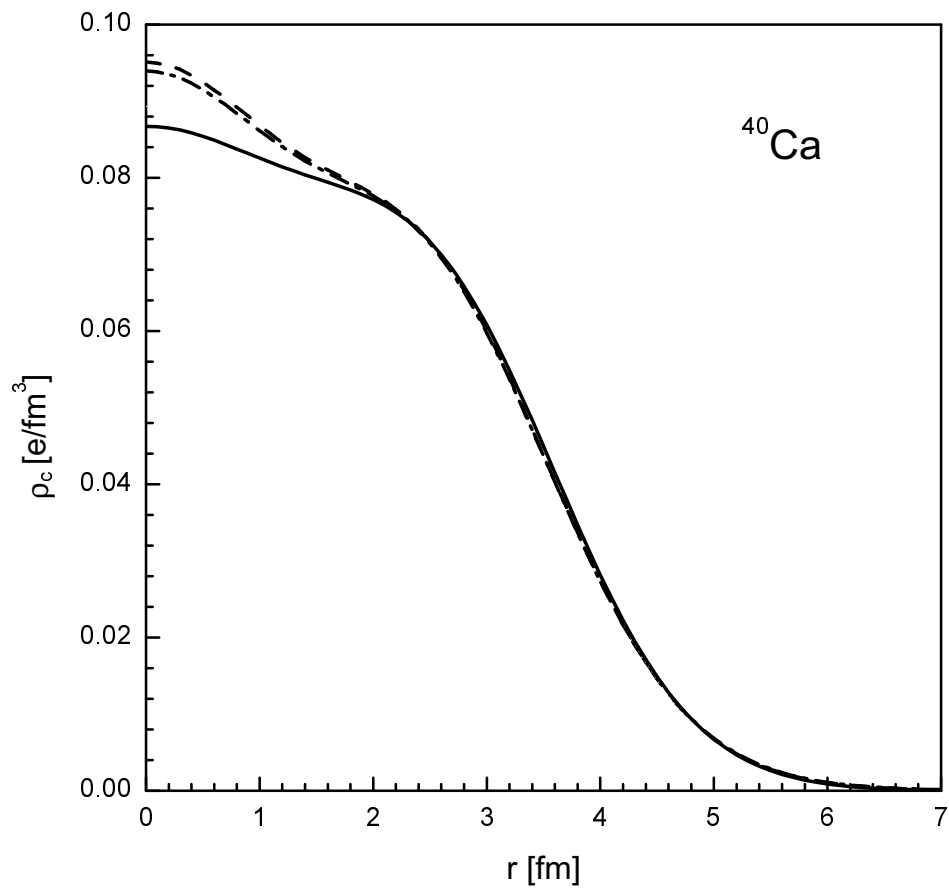
\item[Figure 3:]
The charge density distributions for $^{40}\rm{Ca}$ compared with 
the experimental data (solid curve)~\protect\cite{exp1}. 
The dash-dotted and dashed curves are the
results in the present model with $\chi_c=\frac{1}{2}kr^2$ and
$\chi_c=\frac{1}{2}kr^2(1+\gamma^0)/2$, respectively. 

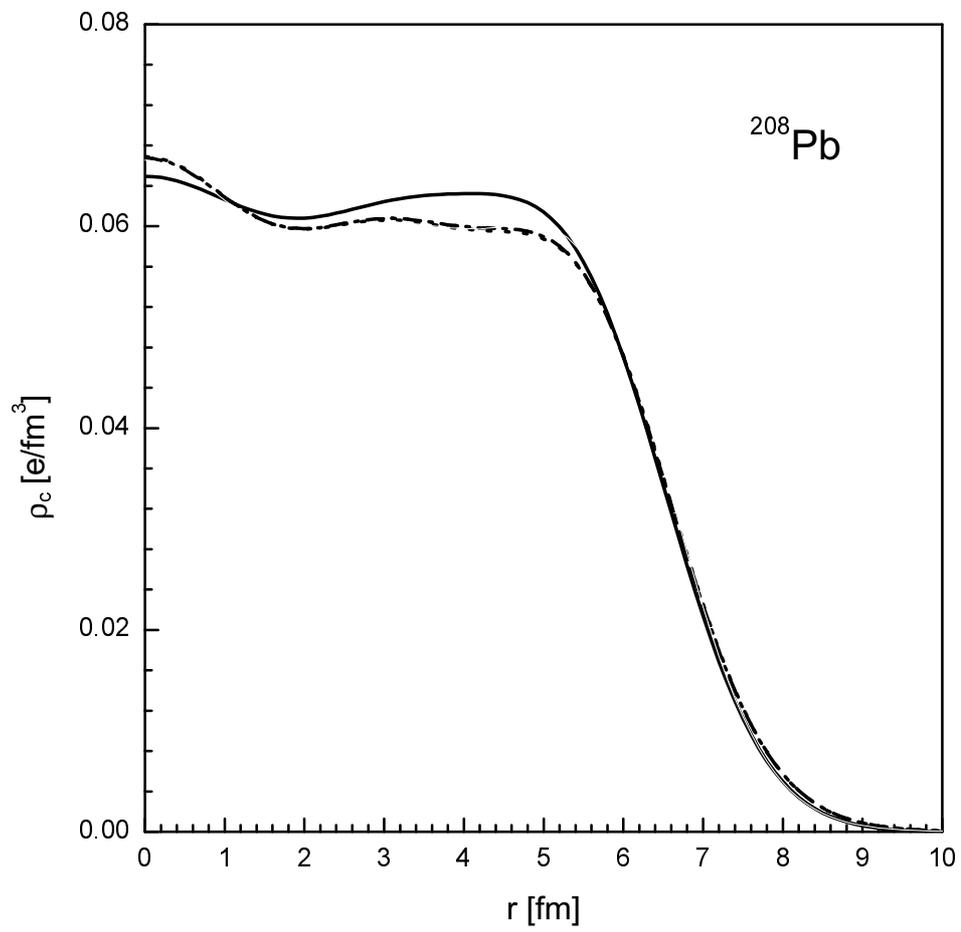
\item[Figure 4:]
Same as Fig. 3 but for $^{208}\rm{Pb}$.

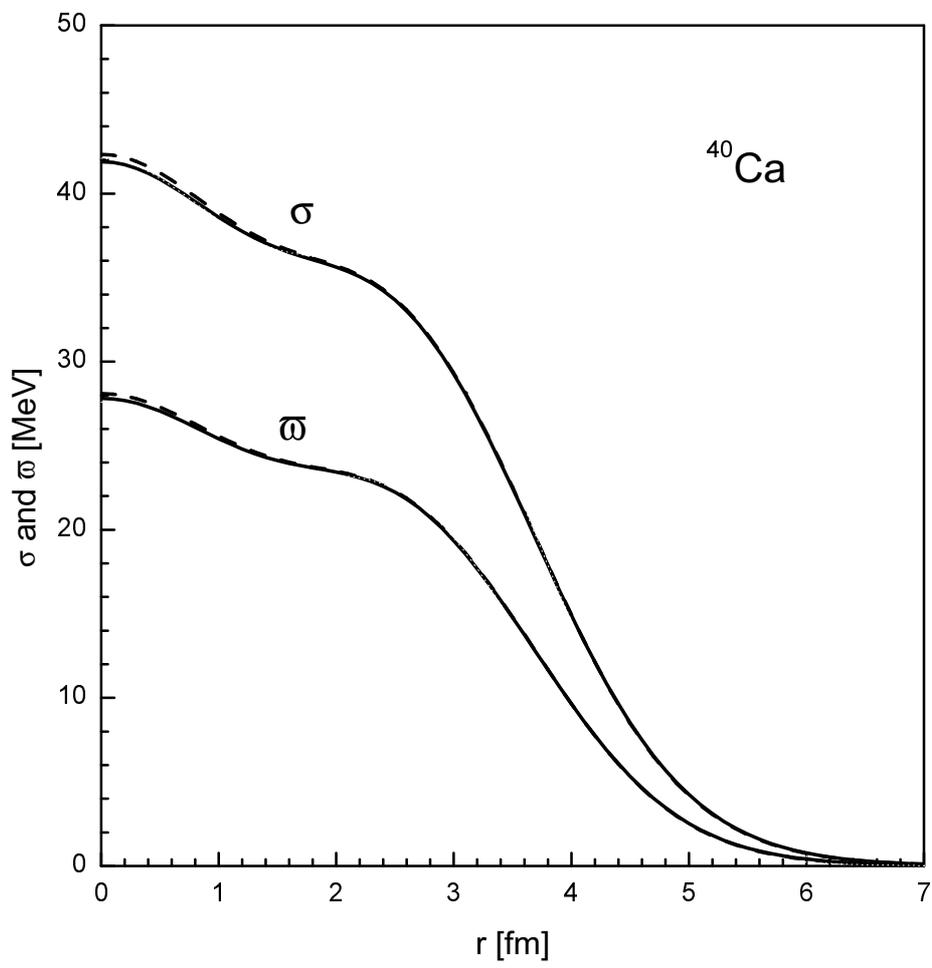
\item[Figure 5:]
The scalar and vector mean fields, $\sigma$ and $\omega$,
as functions of the radial coordinate $r$ for $^{40}\rm{Ca}$.
The results in the present model with $\chi_c=\frac{1}{2}kr^2$ 
are shown by solid curves, while those with
$\chi_c=\frac{1}{2}kr^2(1+\gamma^0)/2$ 
are shown by dashed curves.

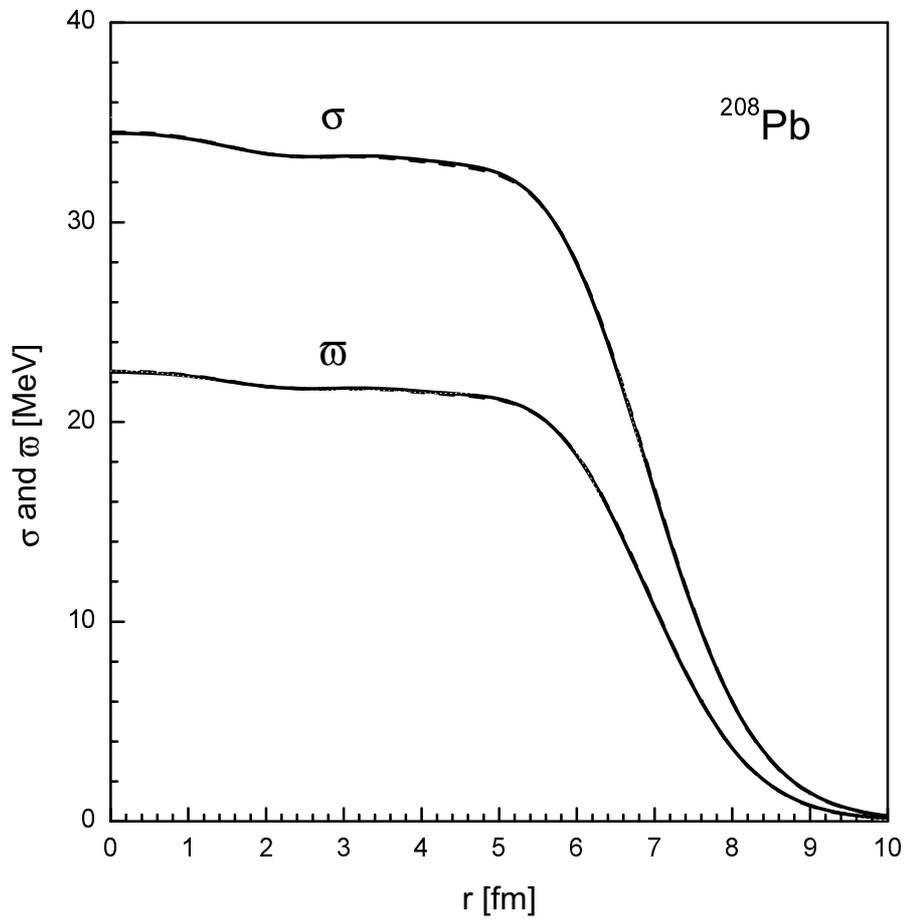
\item[Figure 6:]
Same as Fig. 5 but for $^{208}\rm{Pb}$.

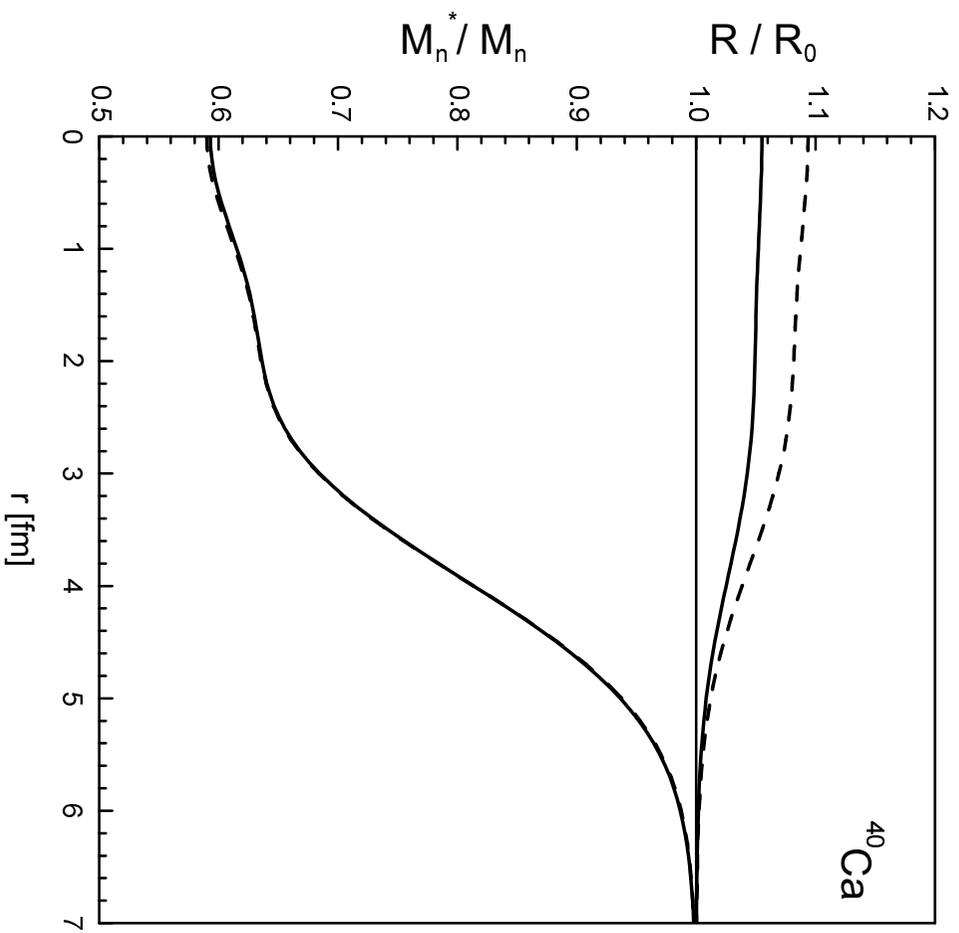
\item[Figure 7:]
The ratios of the nucleon rms radius and effective mass 
in $^{40}\rm{Ca}$ to those in free space, $R/R_0$ and $M^*_n/M_n$,
as functions of the radial coordinate $r$.
The results in the present model with $\chi_c=\frac{1}{2}kr^2$ 
are shown by solid curves, while those with
$\chi_c=\frac{1}{2}kr^2(1+\gamma^0)/2$ 
are shown by dashed curves.

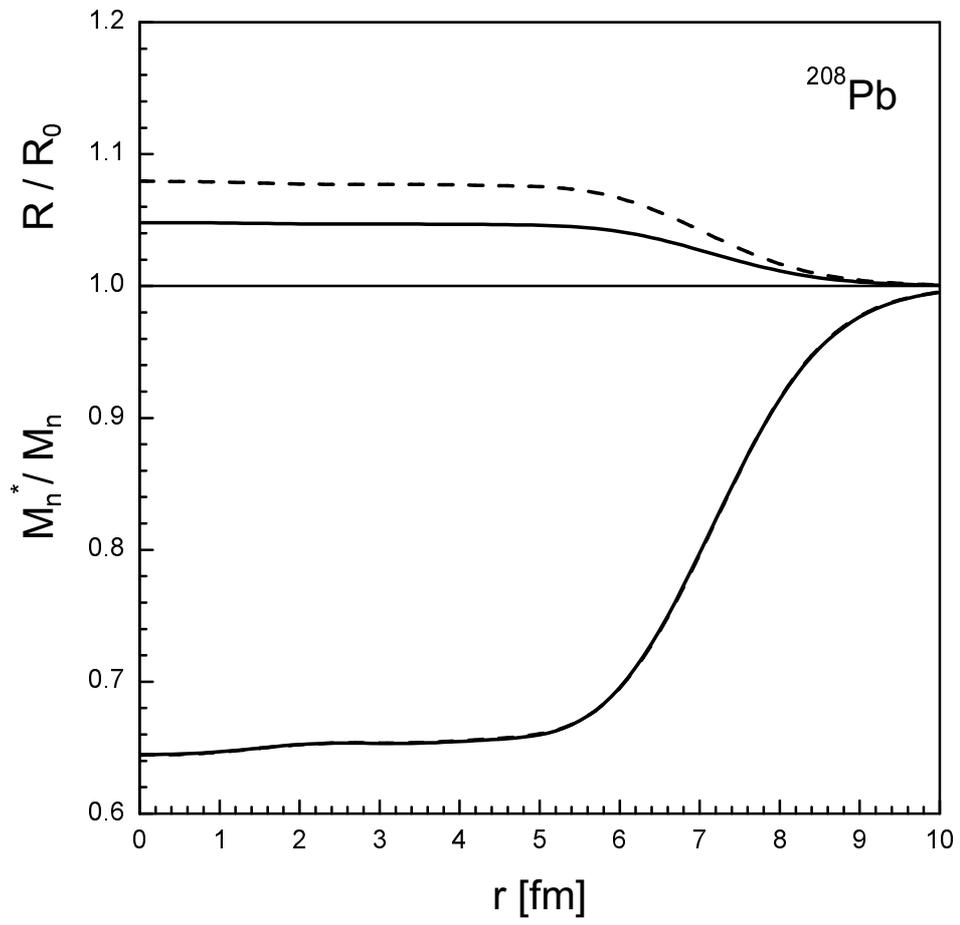
\item[Figure 8:]
Same as Fig. 7 but for $^{208}\rm{Pb}$.

\end{description}

\end{document}